# Empowering Polymeric Materials Discovery by Artificial Intelligence

Chenyao Ma, Linda Zhang, Yuheng Chen, Wei Du, Shangwen Fang, Zihao Jiang, Chuanyu Liu, Xinyu Ma, Rui Su, Gang Wang, Muyao Yu, Dong Zhong, Jie Zhu, Weibo Gong, Huan Gu, Limin Li, Chen Shen, Rui Wu, Zhenghao Wu, Kan Xu, Min Zhou, Donglin He,* Xiayun Huang,* Shan Jiang,* Pengfei Ou,* Jiayu Peng,* Yuwei Zhang,* Jie Zhao,* Di Zhang,* Piao Ma,* Zheng-Hao Li,* and Hao Li*

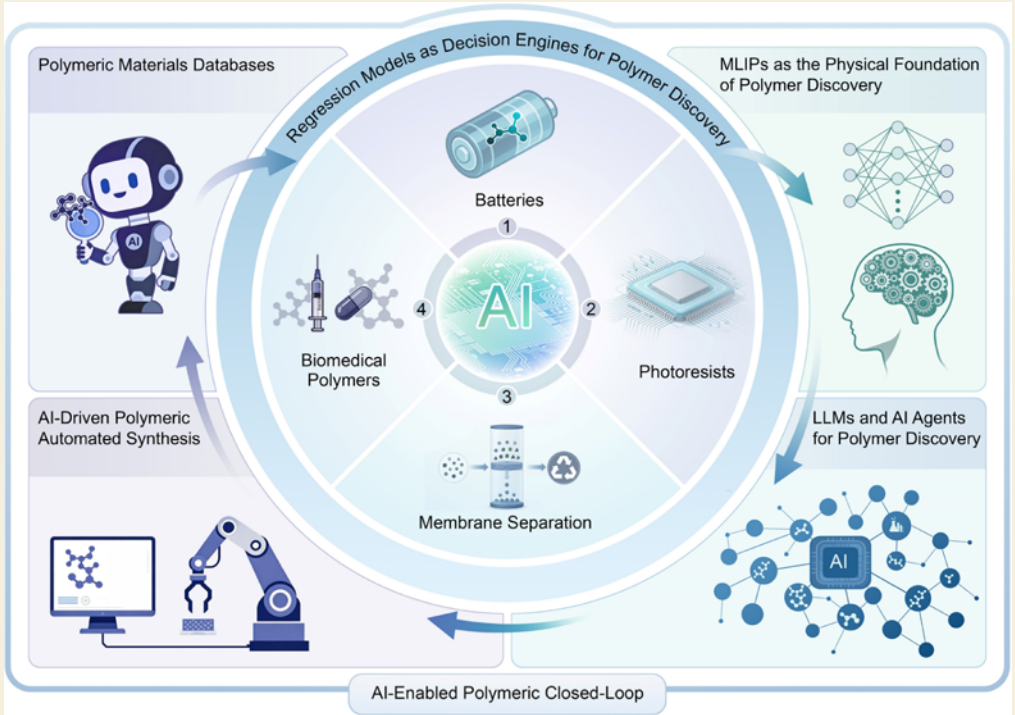

**ABSTRACT:** Polymeric materials underpin modern technologies spanning energy storage, microelectronics, healthcare and sustainable manufacturing. Yet their rational design remains exceptionally challenging because material performance emerges from complex interactions among molecular composition, chain architecture, processing history and hierarchical structural evolution across multiple length and time scales. Consequently, polymer research has long relied on labor-intensive experimentation and fragmented modeling approaches, limiting both mechanistic understanding and innovation efficiency. Recent advances in data infrastructure, machine learning, large artificial intelligence (AI) models and laboratory automation are beginning to reshape this landscape. Rather than functioning as isolated tools, polymer databases, predictive models, AI agents and automated laboratories are increasingly converging into interconnected discovery ecosystems. As a result, the central challenge is shifting from improving predictive accuracy alone to enabling reliable decision-making, adaptive learning and seamless integration across computation, experimentation and scientific reasoning. We argue that polymer science is entering an era of autonomous discovery, in which data, simulation, reasoning and experimentation operate within self-improving feedback loops that continuously generate hypotheses, design materials, execute experiments and refine predictive models. By unifying molecular design, process optimization, experimental validation and industrial translation, such autonomous ecosystems establish a more predictive, reproducible and scalable paradigm for polymer innovation, fundamentally transforming how polymer research is conducted.

## 1. INTRODUCTION

Polymeric materials, macromolecules assembled from repeated monomer units, are indispensable to industrial manufacturing and high-tech applications such as automotive,[1] electronics[2] and healthcare.[3] Representing the mainstream of polymeric materials, global plastic production totals approximately 431 million metric tons in 2025, and is expected to grow to roughly 500 million metric tons by 2030 amid steady expansion of plastics and thermoplastics.[4] Nevertheless, the rapid development of the industry is constrained by inherent material properties.[5] Characterized by complex structures and dynamic behaviors from molecular to macroscopic levels, they are difficult to design rationally, so traditional research still relies on inefficient trial-and-error testing.[6] Against this backdrop, the global manufacturing sector is shifting toward green, low-carbon transformation, imposing stricter requirements on polymer synthesis, processing and end-of-life recycling. Conventional trial-and-error research cannot rapidly develop low-emission, degradable and energy-efficient polymer materials that match carbon neutrality goals, which further amplifies the urgent demand for intelligent, data-driven polymer R&D paradigms.

Emerging tools, including artificial intelligence (AI), multi-scale simulations, and automated laboratories, promise to navigate this complexity.[7] While powerful, these approaches remain constrained by fragmented data sets, inconsistent characterization standards, sparse polymer-specific benchmarks, and weak integration between computation and experiment.[8] Most current strategies focus on individual workflow segments, failing to transform the overall research paradigm.[9] Moving beyond isolated predictive tools, the next leap requires integrated, self-evolving digital polymer ecosystems that translate computational predictions into reliable experimental and engineering decisions.

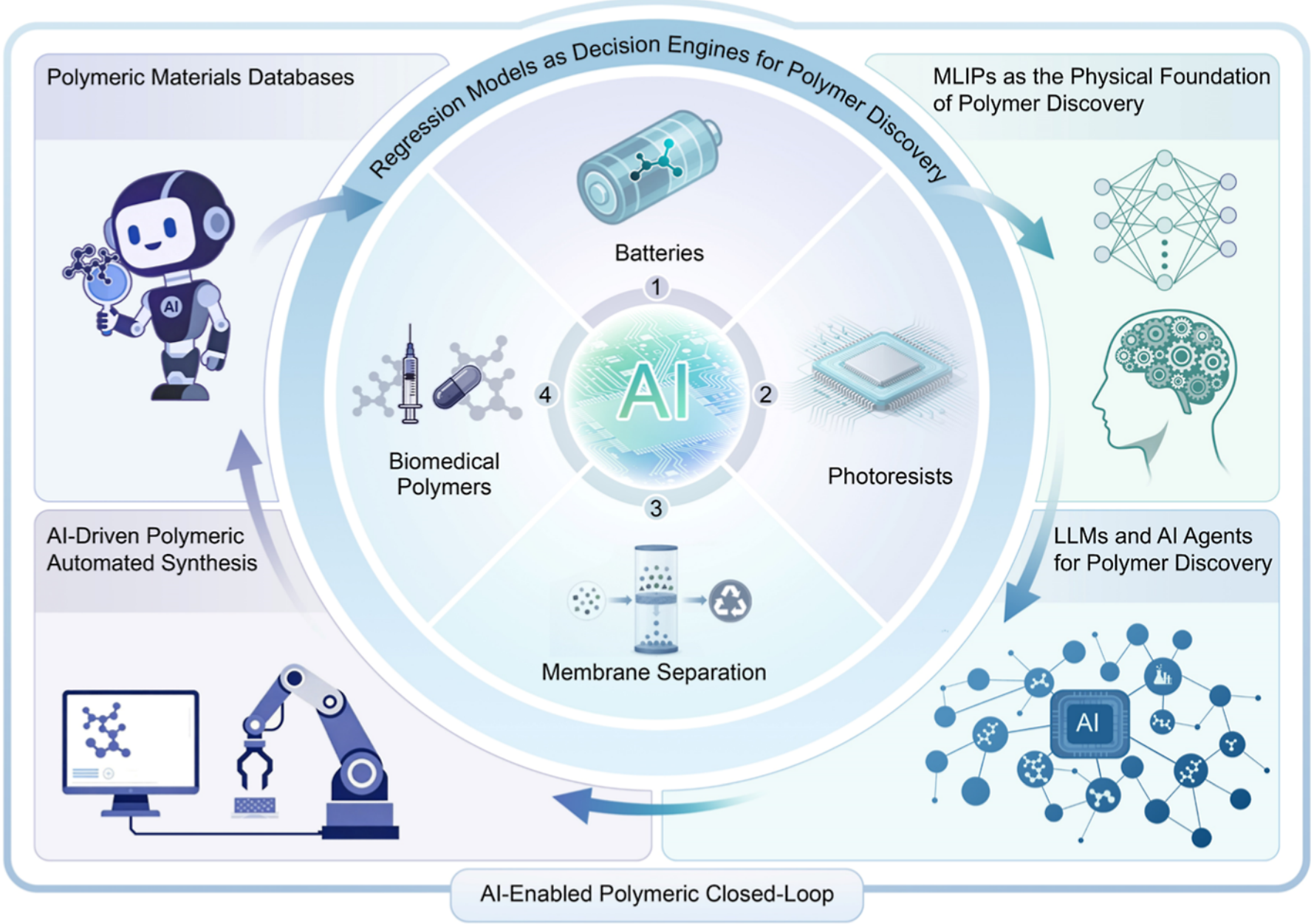

**Figure 1. AI-Enabled Innovation in Polymeric Materials: From Discovery to Closed-Loop Applications.** This framework depicts a hierarchically integrated workflow that underpins data-driven polymer innovation. Starting from the Polymeric Materials Databases, which provide standardized structural, processing, and performance data, regression models serve as decision engines to enable rapid structure−property prediction. MLIPs act as the physical foundation, bridging quantum and mesoscale material behaviors to expand cross-scale predictive capability. LLMs and intelligent agents further orchestrate scientific reasoning, inverse design, and full experimental workflows. The cycle is completed by AI-driven automated synthesis, where polymer candidates are synthesized and characterized, and real-time experimental data is fed back to update the databases and computational models. This iterative loop forms a self-consistent pipeline, enabling continuous self-optimization and accelerating polymer discovery across diverse applications including batteries, photoresists, membrane separation, and biomedical polymers.

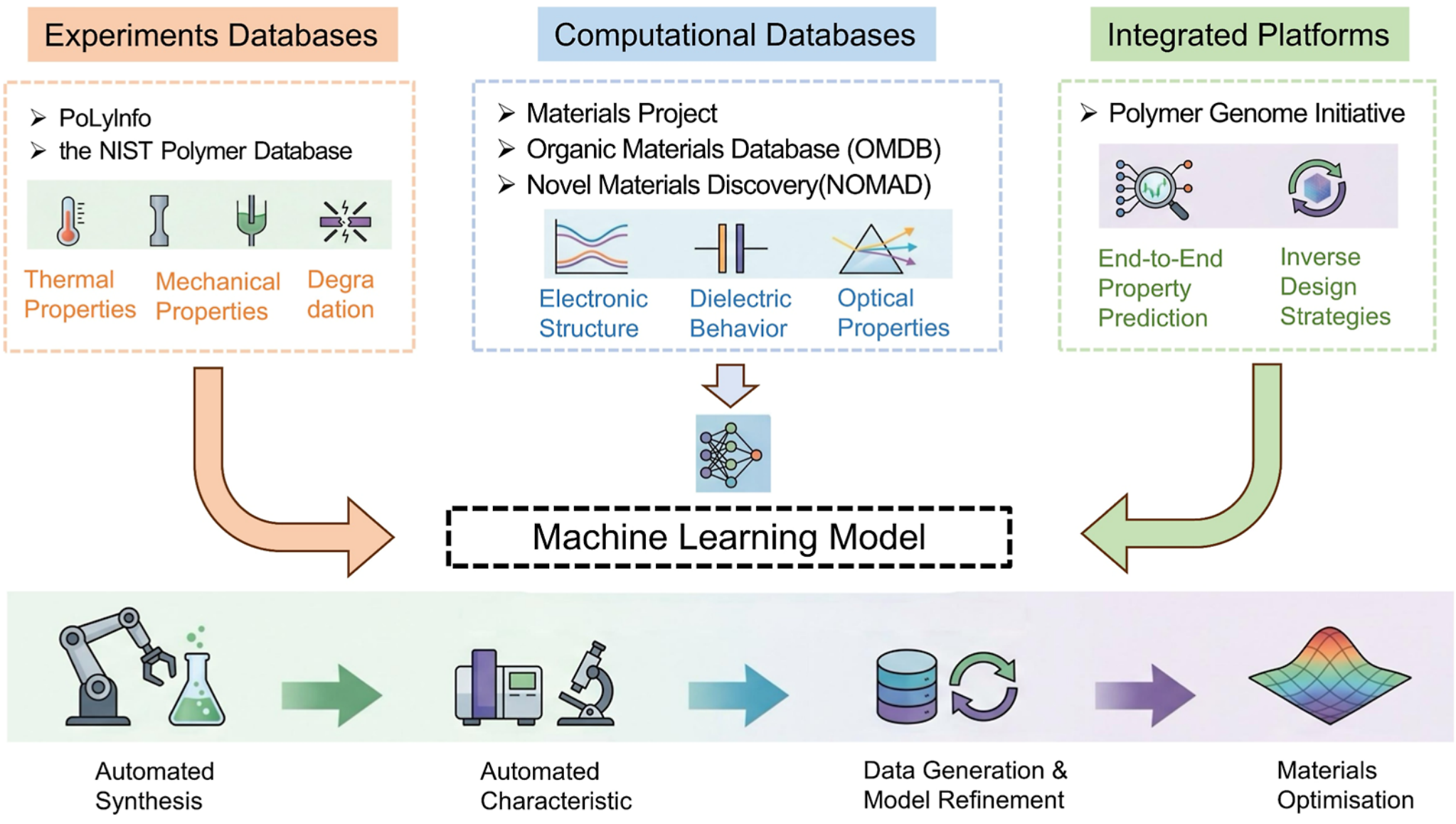

**Figure 2.** Classes of polymeric materials databases and their roles in AI-driven discovery. This schematic illustrates how three categories of polymer databases support machine learning (ML)-driven materials discovery. (a) Experimental databases provide measured data for thermal properties, mechanical properties, and degradation behavior. (b) Computational databases offer calculated data on electronic structure, dielectric behavior, and optical properties. (c) Integrated platforms enable end-to-end property prediction and inverse design strategies.

Herein, we define six core modules that constitute the full-chain intelligent polymer discovery ecosystem: Polymer databases are standardized repositories that curate experimental, simulated and industrial polymer data for long-term data storage and circulation.[10] Regression models refer to data-driven computational tools to establish quantitative structure−performance correlations between microscopic chain structures and macroscopic functional properties of polymers.[11] MLIPs are physics-based simulation potentials fitted from quantum data sets, which enable accurate prediction of atomistic conformational evolution and interfacial interactions of polymer chains.[12] Domain-adapted LLMs are fine-tuned large language models specialized in polymer structure interpretation, literature mining and macromolecular generation.[13] AI agents are task-oriented intelligent executors capable of autonomous decision-making, workflow rearrangement and cross-module interactive iteration.[14] Autonomous laboratories denote robotic closed-loop experimental platforms that accomplish unmanned polymer synthesis, in situ characterization and real-time experimental data backfeeding.[15]

On the basis of the definition and functional division of the above six modules, this *Perspective* discusses how to synergize polymer databases (Figure 1, upper left), regression models, machine learning interatomic potentials (MLIPs) (Figure 1, upper right), LLMs, AI agents (Figure 1, lower right), and autonomous laboratories (Figure 1, lower left) can be synergistically combined to form such ecosystems. Closed-loop feedback across data, simulation, reasoning, and experimentation enables adaptive, scalable, and fully predictive polymer discovery, accelerating the development and real-world deployment of next-generation functional polymers (Figure 1).

## 2. POLYMERIC MATERIALS DATABASES

Reliable and standardized data underpin all AI-powered polymer research. Polymers are governed by complex multiscale interactions among chemical composition, chain architecture, molecular-weight distributions, processing conditions, and hierarchical morphology, making the construction of dedicated data infrastructures substantially more challenging than for many small molecules or inorganic materials.[11] Fragmented data sets, inconsistent reporting practices, and limited interoperability remain major barriers to developing robust and transferable AI models.[16]

Modern polymer databases have evolved into essential infrastructures for predictive modeling, simulation, and autonomous experimentation, and can be broadly categorized into experimental, computational, and integrated platforms.[17] Flagship resources such as PoLyInfo[18] and the NIST Polymer Database (experimental), the Materials Project,[19] Organic Materials Database (OMDB),[20] and Novel Materials Discovery (NOMAD)[21] (computational), and the Polymer Genome Initiative[22] (integrated) collectively provide curated experimental measurements, computational descriptors, and integrated analytical frameworks that support structure−property prediction, inverse design, and large-scale exploration of polymer chemical space. Different databases focus on distinct material features: experimental ones on basic performance, computational ones on microscopic properties, and integrated ones on prediction and inverse design. Beyond serving as repositories, polymer databases are increasingly becoming engines for scientific discovery. For example, large-scale mining of degradation data sets has revealed that polymer solubility can govern degradation pathways, challenging conventional assumptions and demonstrating how data-driven analysis can uncover previously overlooked scientific principles[23,24] (Figure 2).

Most standard polymer structural encoding schemes can describe backbone topology in detail, yet public databases rarely integrate quantitative morphological descriptors such as chain length distribution and cross-linking information into their stored structural entries. Furthermore, polymer systems exhibit prominent stochasticity associated with molecular weight distributions, cross-link density and variable cross-link positions. AI offers promising avenues to develop data-driven, machine-learnable structural descriptors incorporating these distributed morphological characteristics, enabling more realistic encoding of complex heterogeneous polymer architectures.

Although a variety of open-source polymer databases have been established, their data scale and information completeness still cannot support comprehensive polymer materials discovery. Most computational databases rely on static density functional theory (DFT) property calculations, while long-time scale molecular dynamics simulation data sets remain severely scarce. Such data sets are required to predict macroscopic experimental properties including glass transition temperature, melting temperature, viscosity, diffusion coefficients and mechanical viscoelastic responses. To overcome such limitations, the community needs to formulate unified development objectives: expanding data sets covering diverse complex polymer systems, constructing multimodal data resources integrating quantum calculations, molecular dynamics simulations and standardized experimental measurements, and unifying structural representations and metadata standards for polymers.

Despite the continuous expansion of polymer data resources, simply accumulating data cannot eliminate inherent drawbacks of data-driven research. Heterogeneous data quality and insufficient standardization have overtaken limited data volume as the core bottlenecks hindering community progress. The critical challenge lies not only in data quantity, but also in incompatible data representations, the absence of systematic uncertainty quantification, and barriers to cross-scale integration, which greatly restrict the transferability of machine learning models across different polymer systems, research scenarios and experimental platforms. To address these issues, researchers should construct a cross-scale "processing−structure−property−performance" data framework, establish standardized metadata schemas and uncertainty analysis workflows to characterize experimental errors and material heterogeneity. It is also necessary to formulate unified specifications for filtering industrial raw data, noise reduction, missing value imputation and metadata annotation, and build standardized data cleaning pipelines to bridge the gap between laboratory data sets and industrial measurements. In the future, polymer databases will evolve beyond static data repositories toward interconnected dynamic knowledge ecosystems integrating simulation, experimental characterization, predictive modeling and autonomous experimentation. These advanced platforms not only supply data resources but also facilitate iterative learning and rational design, enabling continuous generation, validation and refinement of polymer scientific knowledge. With access to standardized, high-quality and reliable data, regression models can fully unleash their strengths for polymer property prediction and experimental guidance.

## 3. REGRESSION MODELS AS DECISION ENGINES FOR POLYMER DISCOVERY

Among AI methods applied to polymer science, regression models remain one of the most practical and widely adopted approaches owing to their balance of predictive performance, interpretability, and data efficiency.[17,25] Particularly well suited to sparse and heterogeneous polymer data sets, they increasingly support not only property prediction but also materials screening, inverse design, uncertainty quantification, and closed-loop optimization.[26] As their role expands from forecasting properties to guiding decisions, regression frameworks are becoming essential components of autonomous polymer discovery workflows, linking molecular design, processing conditions, and functional performance.[27]

### 3.1. Batteries

Simultaneously maximizing ionic conductivity, electrochemical stability and mechanical durability is a persistent bottleneck for solid polymer electrolytes in advanced batteries.[28] Linked by intricate interactions between polymer structure, salt coordination, chain dynamics, morphology and interfacial characteristics, these performance metrics feature unavoidable trade-offs, making traditional trial-and-error development expensive and inefficient.

At present, material design methodologies based on regression algorithms have demonstrated prominent advantages. On the one hand, they can accurately predict key properties including ionic conductivity,[29] donor number[30] and heat release rate using massive experimental data sets[31] (Figure 3a,b). On the other hand, the combination of regression surrogate models and Bayesian optimization enables rapid exploration of the high-dimensional formulation space and efficient screening of optimal electrolyte systems with high ionic conductivity and a wide electrochemical stability window[32−34] (Figure 3c,d). Beyond property forecasting, these models provide actionable decision guidance to streamline experimental workflows and speed up electrolyte optimization.[35]

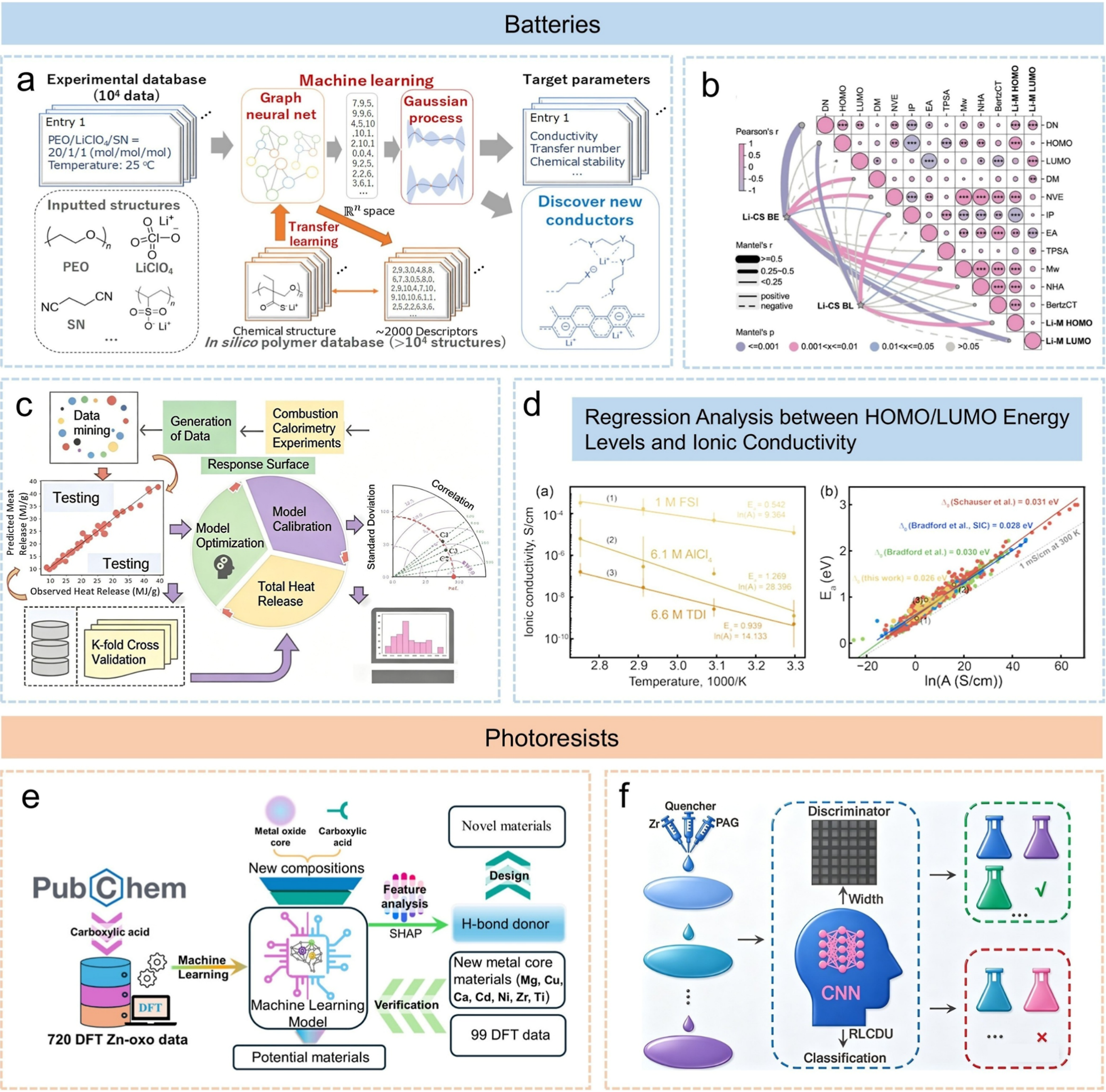

Figure 3. Machine learning-assisted design and regression modeling for polymer electrolytes (batteries) and photoresists. (a) Machine learning workflow for polymer electrolyte design: experimental database construction, graph neural network and Gaussian process modeling, transfer learning, and prediction of ionic conductivity, transference number, and electrochemical stability. (b) Correlation analysis of molecular descriptors and ionic properties for polymer electrolytes. (c) Regression modeling workflow for predicting the total heat release (THR) of polymeric materials, including data mining, model calibration, K-fold cross-validation, and performance testing. (d) Regression analysis of ionic conductivity versus temperature and the correlation between HOMO/LUMO energy levels and ionic conductivity. (e) Machine learning-driven discovery of novel photoresist materials using DFT data and SHAP feature analysis. (f) Convolutional neural network (CNN)-based classification and optimization of photoresist formulations. (a) Reproduced from ref 29. Copyright 2020 American Chemical Society. (b) Reproduced with permission from ref 30. Copyright 2024 Wiley-VCH GmbH. (c) Reproduced with permission from ref 31. Copyright 2025 Elsevier Ltd. (d) Reproduced from ref 33. Available under a CC-BY 4.0 International License. (e) Reproduced from ref 40. Copyright 2024 American Chemical Society. (f) Reproduced from ref 41. Copyright 2024 American Chemical Society.

Nevertheless, existing regression frameworks suffer from inherent drawbacks. Current data sets are dominated by conductivity data, with insufficient records on mechanical performance, interfacial stability, processability and long-term cycling durability. This leads to predicted high-conductivity materials that fail to meet real-world requirements. In addition, the three key performances are usually optimized individually, ignoring their underlying physical correlations. Next-generation regression tools must incorporate physical constraints and uncertainty analysis to combine compositional, processing, interfacial and operational parameters into holistic predictive models. Integrated with active learning, autonomous synthesis and closed-loop pipelines, they will grow into adaptive decision engines to reconcile competing performance demands. Such systems will greatly accelerate industrially viable solid electrolyte development, drastically cut R&D cycles, and lay a solid foundation for fully autonomous battery materials discovery.

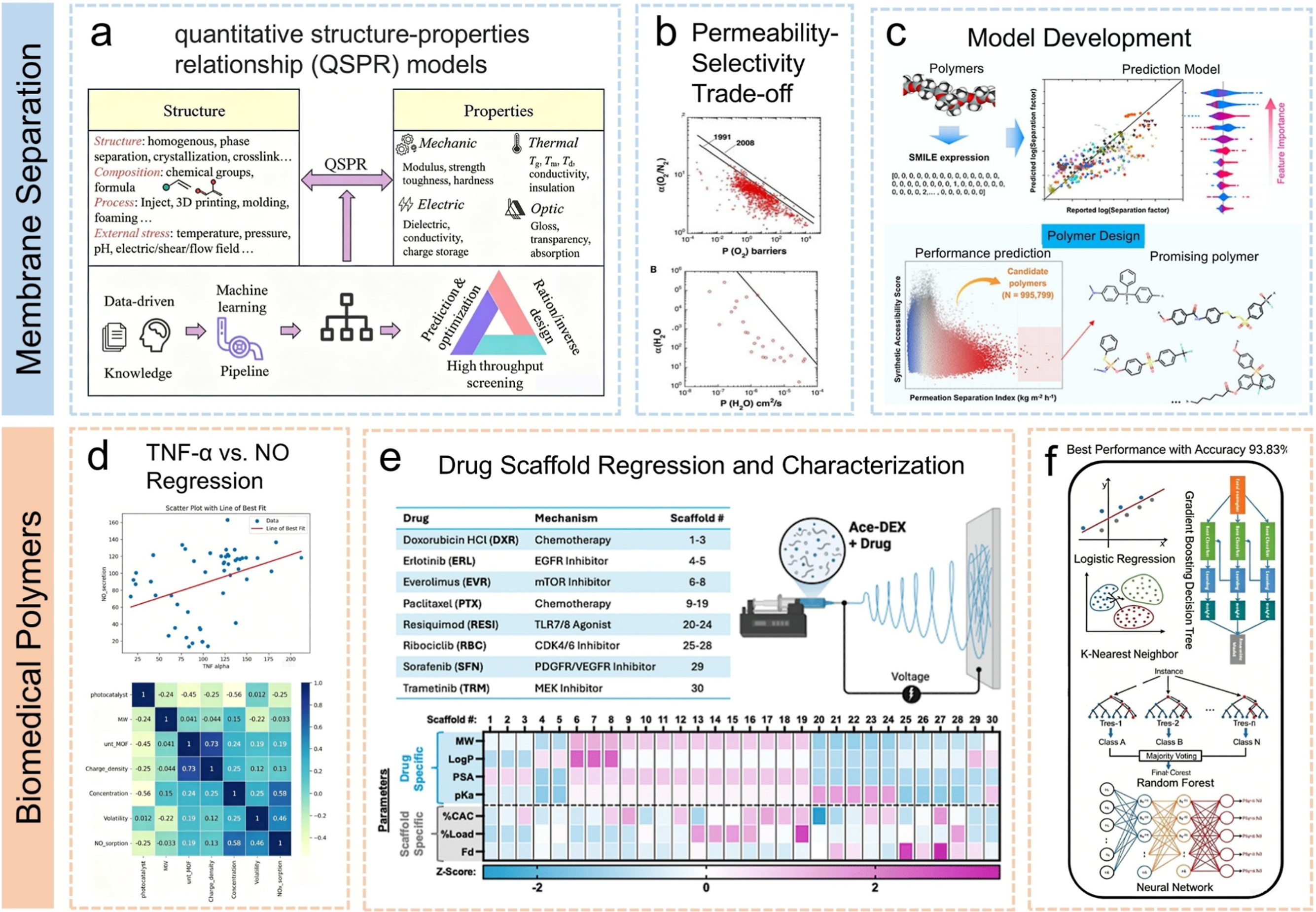

**Figure 4.** Machine learning regression models for polymer materials in membrane separation and biomedical applications. (a) Quantitative structure−property relationship (QSPR) modeling framework for polymers, illustrating the mapping from structural descriptors to functional properties via machine learning pipelines, enabling prediction, optimization, and high-throughput screening. (b) Permeability-selectivity trade-off plots for gas separation membranes, showing the performance distribution relative to the Robeson upper bound. (c) Polymer prediction and design workflow for membrane materials: SMILES-based model development, prediction model validation, and high-throughput screening for promising candidates. (d) Linear regression analysis of TNF-*α* and NO secretion with Pearson correlation matrix for immunomodulatory polymer studies. (e) Drug scaffold regression and characterization: multidrug library, electrospinning fabrication setup, and standardized parameter profiling of Ace-DEX scaffolds. (f) Overview of machine learning algorithms (logistic regression, k-nearest neighbor, gradient boosting decision tree, random forest, and neural network) applied to polymer property prediction, with best classification performance of 93.83%. (a) Reproduced from ref 45. Copyright 2025 The author(s). SmartMat published by Tianjin University and John Wiley & Sons Australia, Ltd. (b) Reproduced from ref 46. Available under a Creative Commons Attribution NonCommercial License 4.0 (CC BY-NC). Copyright 2020 The Authors, some rights reserved; exclusive licensee American Association for the Advancement of Science. No claim to original U.S. Government Works. (c) Reproduced from ref 47. Copyright 2024 American Chemical Society. (d) Reproduced from ref 51. Copyright 2023 The Authors. Advanced NanoBiomed Research published by Wiley-VCH GmbH. (e) Reproduced with permission from ref 52. Copyright 2025 The Royal Society of Chemistry. (f) Reproduced from ref 53. Copyright 2025 The Author(s). Advanced Science published by Wiley-VCH GmbH.

## 3.2. Photoresists

As semiconductor manufacturing advances toward the sub-10 nm regime, overcoming the trade-off among photosensitivity, resolution, and line-edge roughness (LER) has become a defining challenge.[36,37] These metrics result from complex couplings among polymer composition, additives, stochastic photon effects, and process variables such as exposure dose, postexposure bake, developer chemistry, and tool-specific conditions. Consequently, conventional trial-and-error methods struggle to navigate the high-dimensional design space of next-generation lithography, and bridging lab-scale predictions with wafer-scale manufacturing remains a pressing industrial requirement.[38,39]

With high interpretability and practical value, regression models map quantitative composition−process−performance relationships to speed up formulation screening and photo-resist design. Data-driven models including convolutional neural networks, LightGBM and CatBoost are capable of predicting the exposure and dissolution behaviors of chemically amplified resists and uncovering the nonlinear correlations between monomer composition and material performance[40] (Figure 3e). For instance, convolutional neural network models have accurately analyzed the intrinsic links between polymer monomer components and dissolution behaviors[41] (Figure 3f). Gradient boosting algorithms and multiobjective regression methods further expand the exploration scope of chemical systems, realize the simultaneous optimization of sensitivity, critical dimensions and pattern uniformity, and greatly reduce experimental costs and shorten the development cycle.[42]

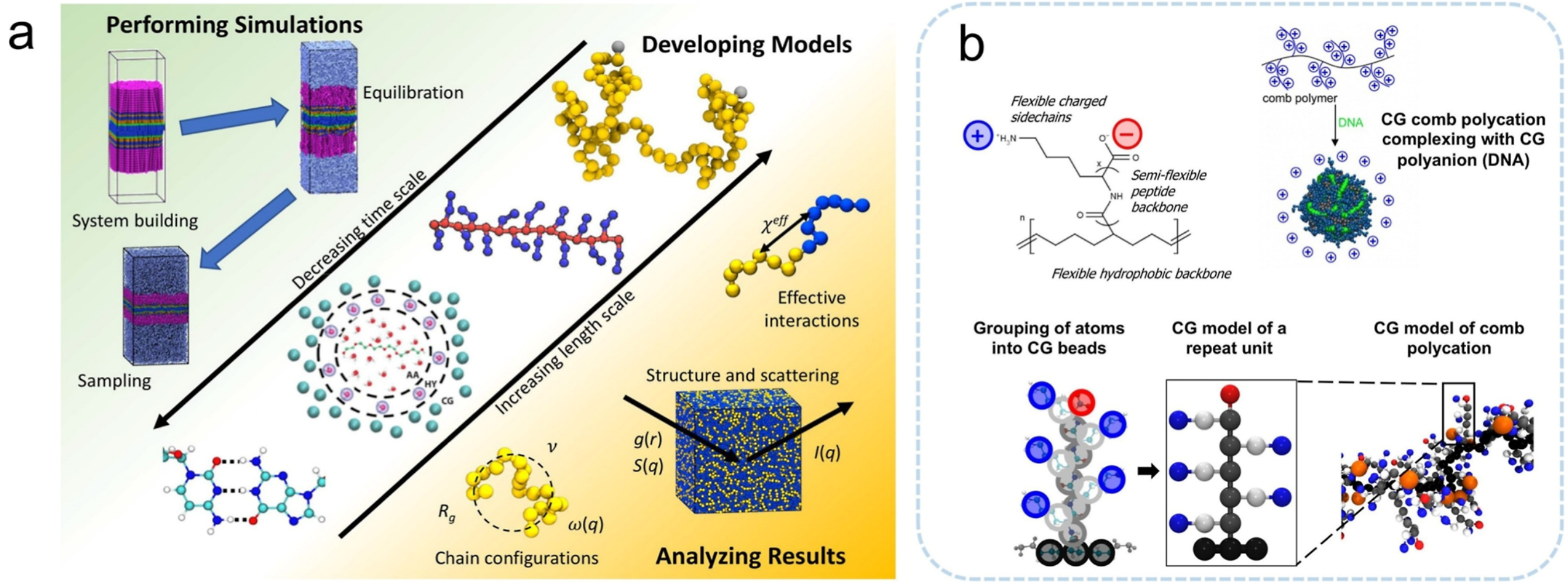

**Figure 5.** Hierarchical framework and benchmarking of MLIPs for polymer property prediction. (a) Overview of core polymer simulation procedures covering model construction, simulation running and result analysis, underpinning the best practices for polymeric modeling. (b) Workflow for developing coarse-grained (CG) models of comb polycations and their DNA complexes, including atom bead grouping, repeat unit definition and polymer assembly. (a,b) Reproduced from ref 54. Copyright 2019 American Chemical Society.

While remarkable progress has been made, substantial challenges still exist. Current models rely heavily on composition-focused lab data sets and inadequately capture process drift, equipment variability, stochastic patterning, and wafer-scale complexities. Moreover, sensitivity, resolution, and LER are often optimized independently, ignoring inherent trade-offs. Future frameworks must integrate physical constraints and uncertainty quantification, unifying molecular descriptors, process parameters, stochastic behaviors, and industrial limits. Combined with Bayesian optimization, active learning, digital twins, and autonomous experimentation, such models can function as adaptive decision engines, co-optimizing materials and manufacturing processes and accelerating the industrial deployment of next-generation photoresists.

## 3.3. Membrane Separation

Polymeric membranes have emerged as leading technologies for gas separation, water purification, and resource recovery owing to their structural tunability, scalability, and low energy requirements.[43] However, their performance is bounded by the Robeson upper bound that creates an unavoidable permeability-selectivity trade-off.[44] Coupled with highly complex (Composition−Processing−Structure−Property−Performance) CPSPP relationships, this barrier greatly limits the output of conventional experimental research. Regression models provide an efficient approach to navigate complex design spaces and quantify structure−performance relationships[45] (Figure 4a).

Regression-based structure−property analysis enables high-throughput virtual screening. It has identified promising candidates breaking the $CO_2/CH_4$ Robeson limit, and multitask or graph-assisted regression can accurately predict multigas transport properties for diverse separation scenarios[46] (Figure 4b). Such tools streamline membrane development by converting molecular information into actionable design principles. Beyond gas separation, regression models also deliver excellent performance in pervaporation-based separation processes. LightGBM models trained on well-curated large data sets achieve a low RMSE of 0.36−0.45 log units[47] (Figure 4c). Such models can screen around $10^6$ hypothetical polymers and efficiently identify candidates with outstanding separation capabilities for organic mixture treatment. Even so, existing data sets are largely confined to ideal laboratory conditions, and their predictive power cannot be reliably extended to full-scale industrial production.

Yet predictive accuracy alone is a poor indicator of industrial utility. Built upon ideal lab data, they fail to capture feed variation, fouling, aging and dynamic operational changes. The community also tends to prioritize permeability and selectivity alone, while neglecting long-term stability and cost competitiveness—key indicators for industrialization. Breaking these bottlenecks demands regression frameworks embedded with multiscale structural information and process characteristics, which combine molecular, morphological and operational descriptors. Together with uncertainty analysis and closed-loop autonomous workflows, these models will grow into intelligent decision systems for joint material and process optimization, accelerating the autonomous discovery of industrially applicable high-performance separation membranes.

## 3.4. Biomedical Polymers

Biomedical polymers underpin drug delivery, implants, tissue engineering and wound healing.[48] Their functionality relies on complex molecular, cellular and physiological interactions in dynamic biological systems. Conventional empirical approaches cannot capture such multiscale dynamics, impeding structure−function analysis and clinical translation.[49,50]

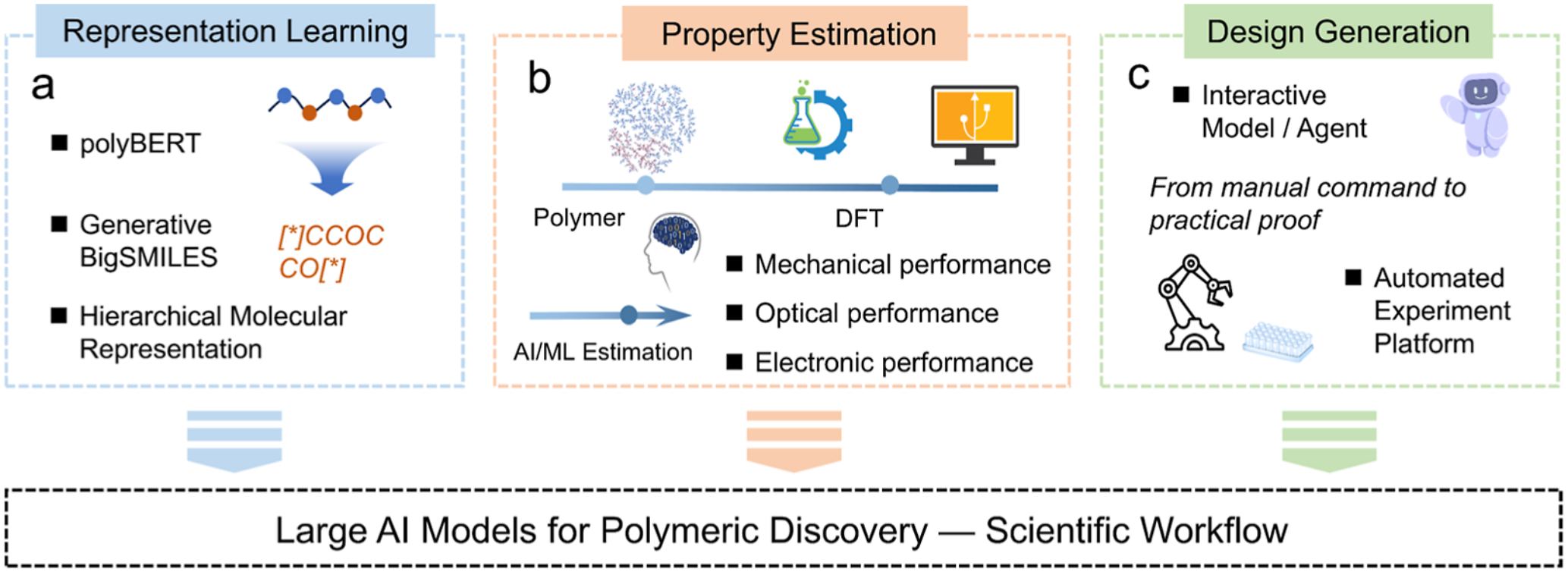

**Figure 6.** Schematic of polymer-oriented foundation models and AI agents for material research. (a) Three typical representation learning approaches for polymers, including PolyBERT, generative BigSMILES with molecular string examples and hierarchical molecular encoding. (b) Two mainstream routes for polymer property evaluation: AI/ML prediction and DFT calculation, which are widely applied to assess mechanical, optical and electronic properties of polymers. (c) The full-cycle workflow of AI agents, covering interactive design generation, automated experimental operation and final result validation, realizing autonomous material research with minimal manual intervention.

Regression models effectively reveal structure-bioactivity links between molecular descriptors and biological behaviors including protein adsorption, drug release and cellular responses[51] (Figure 4d). Gaussian process and ensemble models accurately predict nanoparticle properties and sustained release profiles, enabling mechanism-guided polymer design[52,53] (Figure 4e–f). Though performing well for *in vitro* tests, these tools need integration with patient data, real-time feedback and autonomous optimization to achieve clinically meaningful results.

Predictive accuracy is sometimes not the main limitation; clinical translation poses the core challenge. Current models built on *in vitro* data fail to account for *in vivo* complexities: immune responses, patient heterogeneity, tissue remodeling and degradation effects. Strong laboratory performance therefore rarely translates to reliable clinical outcomes. In addition, existing models focus merely on bioactivity, disregarding manufacturability, reproducibility, regulation compliance and long-term safety. Future cross-scale regression frameworks must unify molecular, cellular, physiological and patient–level parameters. Integrated with uncertainty analysis, digital health data and autonomous workflows, they will become intelligent decision engines for the full pipeline from laboratory discovery to clinical deployment, fostering self-improving systems to advance translational polymer medicine.

## 4. MLIPS AS THE PHYSICAL FOUNDATION OF POLYMER DISCOVERY

MLIPs are emerging as a key technology for bridging a persistent challenge in polymer science: connecting quantum-level chemistry with the length and time scales needed to predict realistic material behavior[54] (Figure 5a,b). Classical force fields can handle large amorphous systems and long trajectories but often fail to capture chemical transferability, polarization, or bond-breaking. Density functional theory (DFT) provides higher accuracy but is computationally prohibitive for realistic polymers. It is worth noting that most MLIPs are developed as viable alternatives to classical force fields for certain polymer systems, instead of serving only as supplementary simulation tools under limited scenarios. Well-trained MLIPs achieve higher accuracy than classical force fields and support both nonreactive and reactive simulations; their capability to describe chemical reactions is determined by the coverage of bond-forming and bond-breaking configurations within training data sets. High costs for data set construction constitute the primary bottleneck for MLIP deployment. Active learning identifies gaps in configurational space and guides targeted DFT calculations, while configurational space visualization enables early detection of missing structures. The combination of these two approaches effectively reduces training data requirements. At present, classical force fields remain irreplaceable for large-scale, long-time scale polymer simulations owing to their low computational cost, well-established parameters and mature software ecosystems. Recent advances in polymer databases and automated simulation infrastructures, such as PoLyInfo[18] and RadonPy,[55] have laid the groundwork for integrating atomistic simulations into AI-driven workflows.

MLIPs are being advanced from system-specific potentials toward fully transferable atomistic models, yet universal generalization across diverse polymer chemistries remains an unresolved challenge. Early neural-network potentials demonstrated that quantum data of short oligomers can be extended to longer polymer chains if local atomic environments are sufficiently sampled. Active-learning strategies enabled robust potentials supporting nanosecond-scale simulations.[56,57] More recent developments, including equivariant graph neural networks (MACE[58]), transferable organic force-field frameworks, and foundation-style atomistic models, broaden applicability across diverse polymer chemistries. These advances point toward simulation models that generalize across polymer classes rather than single systems.

Algorithmic development has advanced swiftly, leaving data infrastructure as the prevailing bottleneck. Most existing MLIPs remain trained on narrow chemical spaces and idealized configurations, limiting transferability across polymer classes, processing histories, and degradation states. Consequently, improvements in benchmark accuracy do not necessarily translate into predictive fidelity for realistic polymer behavior. Workflows such as the PolyData database, Vivace machine learning force fields (MLFF) training, and PolyArena MD simulations exemplify how computational data can be systematically used to develop MLIPs and predict experimental bulk properties like glass transition temperature[59] (Figure 5c). Data sets such as OMol25,[60] OPoly,[61] and PolyData[59] incorporate diverse polymer chemistries, environments, and reactive configurations, while benchmarks like SimPoly[59] shift evaluation toward experimentally relevant observables. Benchmark evaluations, such as those comparing Vivace, MACE-OFF,[62] UMA,[63] and classical force fields (PCFF,[64] OPLS3e[65]) on polymer density prediction, highlight the progress and remaining gaps in transferability[59] (Figure 5d).

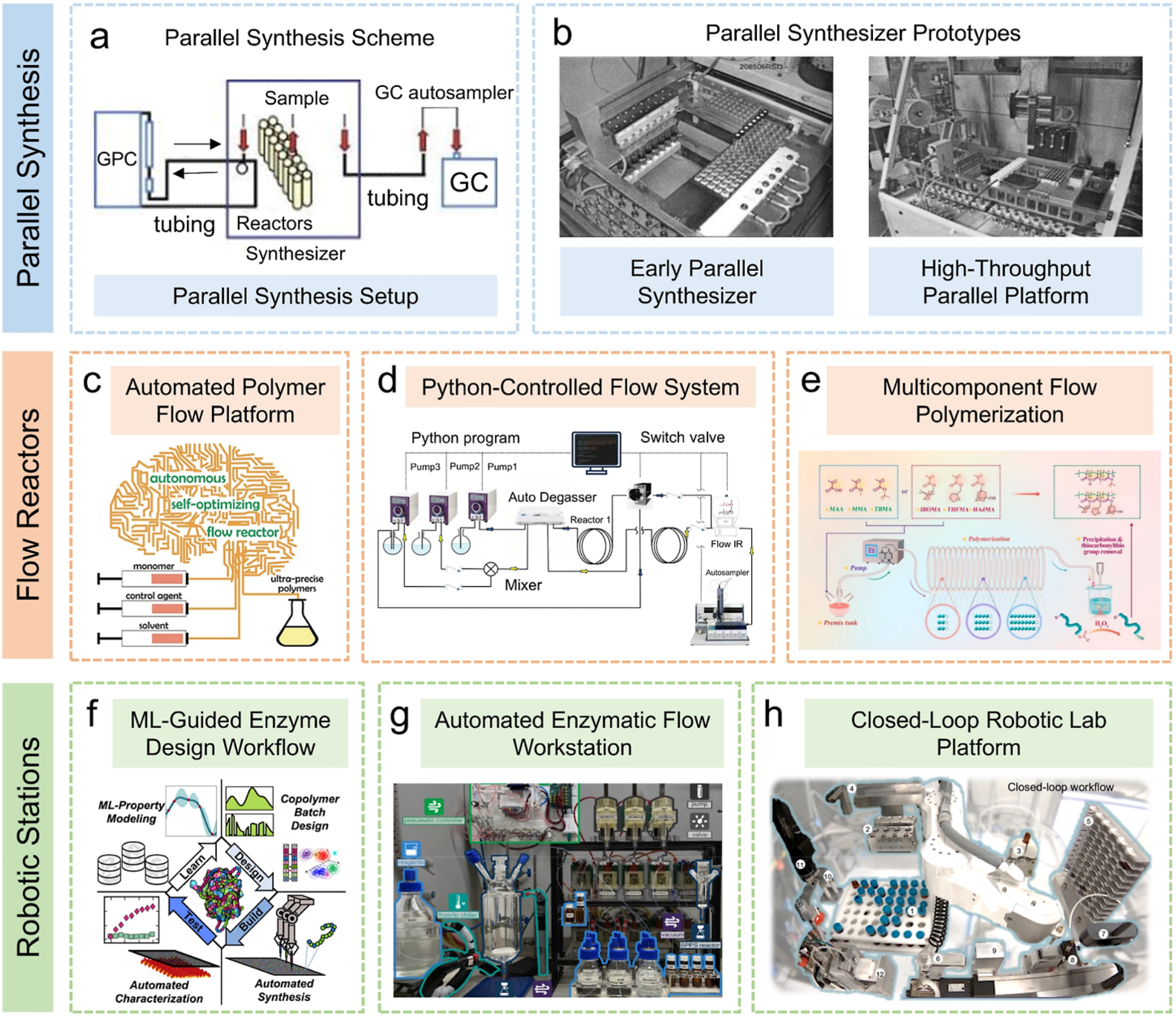

**Figure 7.** Overview of automation technologies for polymer discovery, spanning parallel synthesis, continuous-flow reactors, and closed-loop robotic workstations. (a-b) Parallel synthesis: (a) Typical parallel synthesis setup with reactors, tubing, and integrated GPC/GC autosamplers for high-throughput polymerization and characterization; (b) Prototypes of early parallel synthesizers and high-throughput parallel platforms. (c–e) Continuous-flow reactors: (c) Autonomous self-optimizing polymer flow reactor; (d) Python-controlled flow system with pumps, mixers, and inline IR analysis; (e) Multicomponent flow polymerization with modular reaction steps and product purification. (f–h) Robotic stations: (f) ML-guided iterative workflow combining property modeling, copolymer design, and automated synthesis/characterization; (g) Automated enzymatic flow workstation; (h) Closed-loop robotic lab platform with integrated synthesis, handling, and analysis for self-driving experimentation. (a–b) Reproduced with permission from ref 78. Copyright 2019 AIP Publishing. (c) Reproduced with permission from ref 85. Copyright 2019 Wiley-VCH Verlag GmbH & Co. KGaA, Weinheim. (d) Reproduced with permission from ref 86. Copyright The Royal Society of Chemistry. (e) Reproduced with permission from ref 80. Copyright 2025 Elsevier Ltd. All rights reserved. (f) Reproduced from ref 84. Copyright 2022 The Authors. Advanced Materials published by Wiley-VCH GmbH. (g) Reproduced from ref 82. Copyright 2025 The Author(s). (h) Reproduced with permission from ref 83. Copyright 2025 UChicago Argonne.LLC, and Naisong Shan, Nan Li and Subramanian K.R.S. Sankaranarayanan.

Yet, current MLIP methods, which were mainly developed for mechanism simulations, are usually not targeted for polymer-oriented R&D. They usually struggle with poor extrapolation capacity due to insufficient data for under-explored structure space. Large-scale database, physics-informed MLIP methods for polymers and polymer-specific bechmarking study for MLIPs are urgently needed to make MLIP adapted to polymeric science.

## 5. LLMS AND AI AGENTS FOR POLYMER DISCOVERY

The inherent multiscale hierarchical structures and pronounced processing-dependent properties of polymers pose fundamental challenges to conventional ML methods. Unlike small molecules and inorganic materials, polymers lack a universal molecular representation system, and traditional molecular descriptors fail to fully characterize their stochastic sequences, repeating units, and processing-induced structural variations.[17] This severely limits the capability of data-driven strategies for polymer property prediction and rational design, necessitating the adoption of polymer-specialized foundation models and AI agents to address these core bottlenecks.[66,67]

Foundation models[68] learn transferable representations from large polymer data sets to resolve this issue.[69] Techniques such as PolyBERT,[70] BigSMILES-based[71] language models and chemistry-aware hierarchical encodings generate chemically consistent representations across multiple scales, supporting property prediction, inverse design and cross-domain reasoning (Figure 6a). Nevertheless, today's foundation models are constrained by limited polymer training corpora, weak multimodal integration and static data sets. Since real polymers evolve dynamically throughout synthesis, processing, aging and application, next-generation models need to unify structural, processing, simulation and experimental information via multimodal learning.

Unlike foundation models that focus on knowledge and reasoning, AI agents enable autonomous decision-making and end-to-end workflow management (Figure 6b). They coordinate literature mining, molecular design, simulation and iterative optimization, with representative platforms such as ToPolyAgent and PolyJarvis mainly confined to laboratory-level design tasks without full industrial constraint integration.

Distinct from these lab-only systems, the ArF photoresist AI agent developed by MatSource Tech stands out as a fully industrialized intelligent platform, which delivers full-cycle R&D covering resin molecular design, multicomponent formulation screening, and quantitative industrial performance evaluation under real raw material cost, continuous mass production and regulatory restrictions. While most existing agents are held back by incomplete domain knowledge, poor experimental connectivity and insufficient assessment of manufacturability, cost, regulations and supply chains, thus restricting their use to simplified laboratory settings (Figure 6c). In particular, existing general AI agent frameworks rarely incorporate multiobjective optimization algorithms that simultaneously balance molecular functional performance, raw material expenditure and long-term mass production stability.

Current foundation models and AI agents remain far from mature for polymer-oriented R&D. Foundation models struggle with dynamic material behaviors and cross-scale feature learning, whereas AI agents still exhibit limited adaptability to industrial constraints and uncertain experimental environments. Overcoming these barriers will require polymer-specific foundation data sets, stronger multimodal learning frameworks, improved physical interpretability, and deeper integration between digital intelligence and laboratory infrastructure. More importantly, future progress should focus not only on improving model performance but also on enabling seamless interactions among knowledge extraction,[72] simulation,[73] experimentation,[74] and manufacturing. In such a framework, foundation models may serve as cognitive engines for scientific reasoning, while AI agents coordinate decision-making and execution across the entire discovery pipeline. Together, they could form the core of autonomous polymer discovery ecosystems capable of continuously generating hypotheses, designing materials, performing experiments, and refining knowledge with minimal human intervention.

## 6. AI-DRIVEN POLYMERIC AUTOMATED SYNTHESIS

Predictive models alone cannot achieve fully autonomous polymer discovery. Progress is hindered by the difficulty of generating high-quality, reproducible experimental data to validate models, explore untapped polymer design spaces, and optimize performance. Laboratory automation provides the essential experimental counterpart, bridging digital predictions and physical validation, and enabling iterative feedback that transforms fragmented experiments into closed-loop discovery workflows.[75,76]

First-generation automation focused on parallel synthesis to increase throughput.[77] Conventional polymer synthesis suffers from low efficiency, constraining large-scale screening of compositional and processing parameters. Parallel platforms, such as Chemspeed ASW2000 and SLT1000, integrate environmental control and automated handling to accelerate polymerization, characterize molecular weight distributions, and construct standardized data sets that feed into ML models[78] (Figure 7a,b). Using automation and intelligent optimization technologies, researchers optimized the spray-coating process for iridescent hydroxypropyl cellulose thin films, clarified the optical mechanism and verified its application prospects, providing a reference for the automated preparation of thin films.[79]

Modern systems advance beyond high-throughput screening toward real-time optimization and autonomous decision-making. Continuous-flow polymerization[80] achieves precise reaction control and inline analysis, while robotic stations and self-driving laboratories, including platforms like Chemputer[81,82] and Polybot,[83] integrate synthesis, characterization, data management, and computational optimization into unified pipelines. A copolymer design strategy based on active machine learning, coupled with automated synthesis and characterization platforms, has identified polymers that preserve or even enhance enzyme activity under thermal denaturation conditions. Machine learning guides iterative experimentation to build adaptive workflows for co-optimizing materials and processes, offering a new path to custom polymer−protein hybrids and broader materials design applications.[84] ML guides iterative experimentation, enabling adaptive workflows that optimize both materials and processes[85−87] (Figure 7c−h).

Several critical bottlenecks still limit the advancement of automated polymer synthesis. For cross-linked systems, unattended automation suffers from uneven cross-linker feeding, absent in situ gelation monitoring, and pipeline clogging by continuously formed cured gels. For high-viscosity melts and concentrated prepolymer solutions, automated transportation, online sampling and real-time viscosity control remain challenging, destabilizing closed-loop experimental workflows. For oxidation-prone monomers sensitive to oxygen and moisture, fully automated pipelines require continuous high-purity inert protection, precision closed feeding and online oxygen detection, which are rarely integrated in current robotic synthesis platforms. Moreover, existing automation platforms poorly couple with diverse AI models, show low compatibility with multiscale and complex chemical systems, and lack scalable industrial potential. Future efforts should integrate automated systems with multimodal data, uncertainty quantification and manufacturing-adaptive AI agents. Integrated closed-loop frameworks combining predictive models, MLIPs, regression tools and autonomous laboratories can build self-evolving polymer research pipelines, accelerate material development, shorten research cycles, and bridge the gap between fundamental laboratory studies and industrial applications.

## 7. AI-ENABLED POLYMERIC CLOSED-LOOP

Autonomous polymer discovery is underpinned by a hierarchically integrated closed-loop ecosystem coupling computational modules and automated experiments.[14] Standardized multiscale polymer databases provide fundamental structural, processing and performance data. Regression models enable rapid structure−property prediction, while MLIPs bridge quantum and mesoscale material behaviors to expand cross-scale predictive capability. LLMs and AI agents further orchestrate scientific reasoning, inverse design and full experimental workflows. Automated laboratories synthesize and characterize polymer candidates, feeding real-time experimental data back to update computational models. This iterative loop forms a self-consistent pipeline for autonomous polymer discovery.[77,88]

Closed-loop systems drive a pivotal shift of polymer research from empirical trial-and-error to data-driven innovation. Conventional studies are limited by fragmented experiments and isolated computations, featuring low data efficiency, long cycles and restricted parameter exploration. In contrast, closed-loop workflows transform static predictions into dynamic active learning. Continuous experimental feedback iteratively optimizes models, material designs and processing conditions, endowing the system with self-evolving capacity.[83,89] This integrated paradigm greatly broadens polymer design spaces and accelerates efficient material innovation[83] (Figure 8).

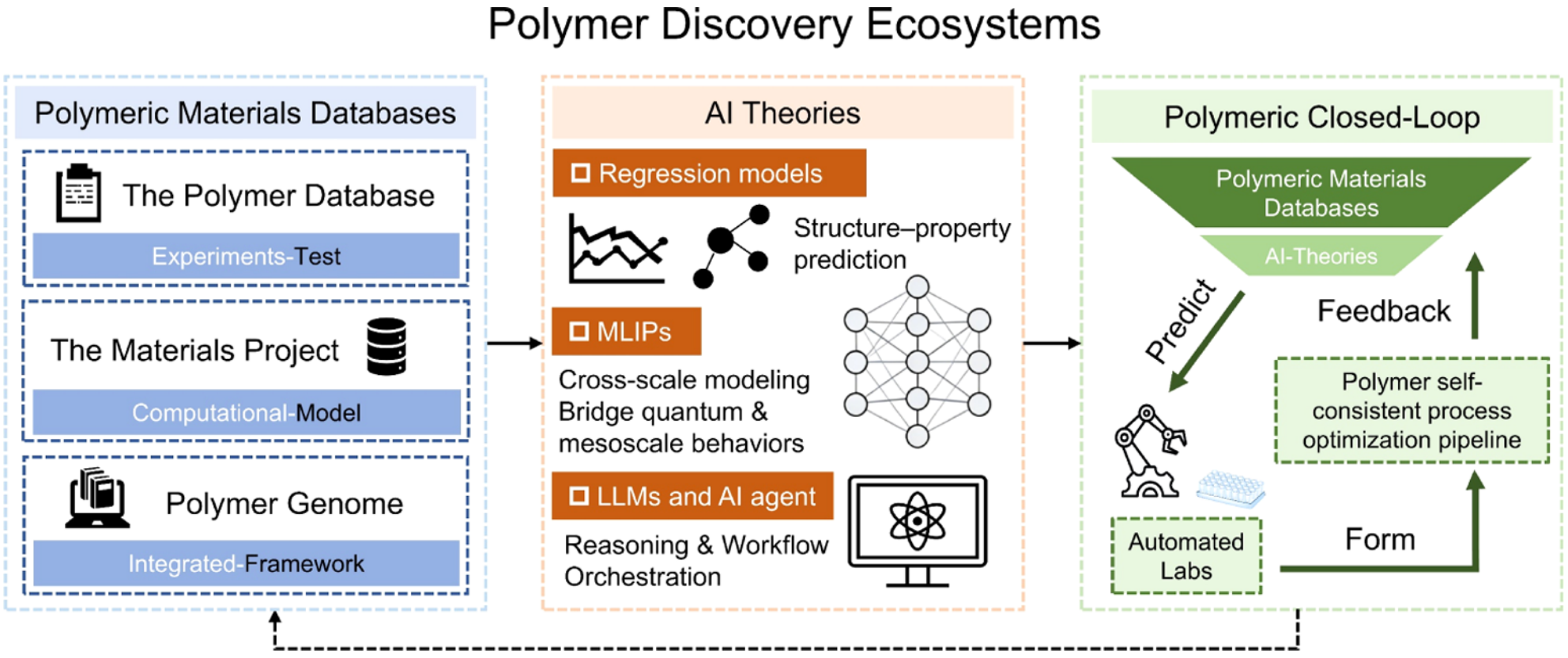

**Figure 8.** Schematic of the autonomous polymer discovery ecosystem. This three-module framework illustrates the closed-loop pipeline for polymer innovation. The Polymeric Materials Databases module (left) integrates experimental, computational, and multisource framework data, represented by PolyInfo, The Materials Project, and Polymer Genome. The AI Theories module (middle) leverages regression models for structure−property prediction, MLIPs to bridge quantum and mesoscale behaviors, and LLMs/AI agents for scientific reasoning and workflow orchestration. The Polymeric Closed-Loop module (right) forms a self-consistent process optimization pipeline: AI-derived predictions guide automated laboratory synthesis and characterization, whose experimental data is fed back to update the databases and models, enabling iterative self-optimization.

Although the vision of autonomous polymer discovery is increasingly tangible, significant challenges remain before closed-loop systems can operate reliably at scale. Poor data standardization, limited multimodal integration, unquantified prediction uncertainty, and weak interoperability between computational modules and laboratory platforms continue to restrict overall performance. Moreover, most existing workflows are optimized for academic demonstration rather than industrial deployment, often neglecting manufacturability, cost, sustainability, and regulatory constraints. As a result, most current closed-loop platforms remain proof-of-concept demonstrations rather than deployable industrial discovery systems. Addressing these bottlenecks will require harmonized data infrastructures, robust multimodal learning, uncertainty-aware decision frameworks, and tighter integration of automation with real-world operational constraints—steps that are essential to fully realize the potential of self-optimizing, autonomous polymer discovery ecosystems.

## 8. SUMMARY AND OUTLOOK

Modern polymer research is evolving rapidly from empirical trial-and-error toward a digital, intelligent, and autonomous innovation paradigm. Driven by multiscale material data accumulation, machine learning algorithms, and LLM-based intelligent agents, polymer R&D is shifting from passive verification of known rules to active exploration of unknown material spaces. The integration of computational prediction, automated experimentation, and intelligent decision-making has become the dominant trend, fundamentally upgrading traditional discrete research modes and laying a solid foundation for accelerated iteration of high-performance and functional polymers.

Nevertheless, a number of prominent challenges have emerged in the process of technology implementation and system improvement, which can be mainly categorized into six major aspects:

(i) Fragmented and noncyclic polymer database construction. Current polymer databases adopt inconsistent data annotation and collection criteria, with scattered experimental, simulation and industrial data sets. Most databases only support one-way data storage, lacking automated data update channels fed by simulation outputs and experimental results, which destroys the basic data circulation loop of the whole ecosystem.

(ii) Absent physical mechanism models for fundamental polymer structure−performance correlation deciphering. Current data-centric polymer AI systems lack universal physical models to elaborate intrinsic structure−property relationships. This deficiency not only limits the predictive accuracy and generalization capability of machine learning models, but also imposes insufficient physical constraints on generative model outputs, ultimately resulting in poor model interpretability and unreliable material design results.

(iii) Decoupled iteration between regression models and MLIPs. Empirical regression models fail to build scalable structure−property correlations for amorphous and cross-linked polymers, while MLIPs usually suffer from poor generalization for complex polymer conformational evolution. The two simulation tools lack bidirectional parameter tuning and result mutual correction, failing to form an internal simulation-data closed loop for multiscale polymer prediction.

(iv) Imperfect autonomous reasoning loop of LLMs and AI agents. Generic LLMs lack polymer-specific domain adaptation, and existing AI agents cannot realize autonomous workflow iteration. Specifically, many current agents are unable to connect data interpretation, model revision, experimental scheme redesign and result evaluation, breaking the intelligent reasoning closed loop within digital decision-making units.

(v) One-way experimental operation of autonomous high-throughput laboratories. Current autonomous laboratory platforms can only finish forward synthesis and characterization tasks for lab-scale polymer samples. They have no built-in data backfeeding interfaces, so measured performance data cannot be automatically uploaded to databases or fed back to modeling tools, resulting in open-loop experimental workflows without iterative optimization.

(vi) Disrupted global data-simulation-reasoning-experiment closed-loop ecosystem. There is no unified middleware to realize information intercommunication among databases, regression models, MLIPs, AI models, agents and experimental platforms. The isolated standalone modules cannot form collaborative cyclic operation. Furthermore, industrial manufacturability, cost and sustainability indicators are excluded from the whole closed-loop workflow, eventually leading to academic prototypes difficult to be deployed in polymer industrial scenarios including batteries, photoresists, membranes, and biomedical polymers.

Future polymer innovation requires hierarchical, self-evolving digital materials ecosystems to address the aforementioned six critical challenges and build a full-cycle polymer research paradigm. First, standardized, updatable multimodal databases integrating molecular structures, processing parameters, performance metrics, degradation pathways and industrial constraints will unify data criteria and realize automatic data circulation, solving database fragmentation and one-way data storage issues. Second, Advancing multiscale physical exploration combining mesoscopic and coarse-grained simulations to derive physically interpretable constitutive formulas. Beyond mechanism-driven physical modeling, symbolic regression embedded within standardized polymer databases is encouraged to screen latent microstructure descriptors and establish physics-consistent governing equations. Notably, such data-driven equation mining paradigm remains vastly underexplored in polymer science, while it has achieved proven feasible and instructive success in allied material fields typified by hydrogen storage materials.[90,91] Third, coupled regression models and MLIPs will achieve bidirectional parameter correction and cross-scale iteration, eliminating model decoupling and poor generalization limitations. Fourth, domain-customized LLMs and interactive AI agents will complete end-to-end autonomous reasoning, scheme redesign and result evaluation, optimizing the internal intelligent decision loop. Fifth, closed-loop upgraded autonomous laboratories will feed experimental data back to modeling modules and databases, transforming open-loop experimental testing into iterative experimental optimization. Sixth, cross-module information interconnection will construct intact data-simulation-reasoning-experiment global closed-loop ecosystems, bridging the gap between academic design and industrial manufacturability, cost and sustainability demands. Such optimized ecosystems will expand polymer design spaces, mitigate inherent performance trade-offs, and accelerate the development of industrialized high-performance, cost-effective, and sustainable polymers.

Overall, this *Perspective* clarifies module-level defects and closed-loop barriers of current polymer intelligent R&D, and provides targeted construction routes for next-generation autonomous polymer discovery ecosystems, guiding the future industrial-oriented digital innovation of functional polymers.

## ■ AUTHOR INFORMATION

### Corresponding Authors

**Donglin He** − *Thrust of Sustainable Energy and Environment, The Hong Kong University of Science and Technology (Guangzhou), Guangzhou 511453 Guangdong, China;* orcid.org/0000-0002-5512-8162; Email: donglinhe@hkust-gz.edu.cn

**Xiayun Huang** − *The State Key Laboratory of Molecular Engineering of Polymers and Department of Macromolecular Science, Fudan University, Shanghai 200438, P. R. China;* orcid.org/0000-0003-3053-7354; Email: huangxiayun@fudan.edu.cn

**Shan Jiang** − *School of Physical Science and Technology, ShanghaiTech University, Shanghai 201210, P. R. China;* orcid.org/0000-0002-3990-8687; Email: jiangshan@shanghaitech.edu.cn

**Pengfei Ou** − *Department of Chemistry, National University of Singapore, Singapore 117543, Singapore;* orcid.org/0000-0002-3630-0385; Email: pengf.ou@nus.edu.sg

**Jiayu Peng** − *Department of Materials Design and Innovation, University at Buffalo, Buffalo, New York 14260, United States*; Email: jypeng@buffalo.edu

**Yuwei Zhang** − *Jiangsu Key Laboratory of New Power Batteries, Jiangsu Collaborative Innovation Centre of Biomedical Functional Materials, School of Chemistry and Materials Science, Nanjing Normal University, Nanjing 210023, P. R. China;* orcid.org/0009-0002-8741-3889; Email: ywzhang@nnu.edu.cn

**Jie Zhao** − *College of Smart Materials and Future Energy, State Key Laboratory of Molecular Engineering of Polymers, Fudan University, Shanghai 200433, China;* orcid.org/0000-0003-4543-3494; Email: jiezhao@fudan.edu.cn

**Di Zhang** − *Advanced Institute for Materials Research (WPI-AIMR) and Frontier Research Institute for Interdisciplinary Sciences (FRIS), Tohoku University, Sendai 980-8577, Japan;* orcid.org/0000-0001-6347-9344; Email: di.zhang.a8@tohoku.ac.jp

**Piao Ma** − *Suzhou MatSource Technology Co., Ltd., Suzhou 215000 Jiangsu, China; Gusu Laboratory of Materials, Suzhou 215000 Jiangsu, China*; Email: mapiao@matsourceai.com

**Zheng-Hao Li** − *State Key Laboratory of Advanced Environmental Technology, School of Environment, University of Science and Technology of China, Hefei 230026, China*; Email: lizhhao@ustc.edu.cn

**Hao Li** − *Advanced Institute for Materials Research (WPI-AIMR), Tohoku University, Sendai 980-8577, Japan;* orcid.org/0000-0002-7577-1366; Email: li.hao.b8@tohoku.ac.jp

### Authors

**Chenyao Ma** − *Suzhou MatSource Technology Co., Ltd., Suzhou 215000 Jiangsu, China*

**Linda Zhang** − *Advanced Institute for Materials Research (WPI-AIMR) and Frontier Research Institute for Interdisciplinary Sciences (FRIS), Tohoku University, Sendai 980-8577, Japan*

**Yuheng Chen** − *Jiangsu Key Laboratory of New Power Batteries, Jiangsu Collaborative Innovation Centre of Biomedical Functional Materials, School of Chemistry and Materials Science, Nanjing Normal University, Nanjing 210023, P. R. China*

**Wei Du** − *Suzhou MatSource Technology Co., Ltd., Suzhou 215000 Jiangsu, China; Gusu Laboratory of Materials, Suzhou 215000 Jiangsu, China*

**Shangwen Fang** − *Department of Chemistry, National University of Singapore, Singapore 117543, Singapore*

**Zihao Jiang** − *Thrust of Sustainable Energy and Environment, The Hong Kong University of Science and Technology (Guangzhou), Guangzhou 511453 Guangdong, China*
**Chuanyu Liu** − *Department of Materials Design and Innovation, University at Buffalo, Buffalo, New York 14260, United States*
**Xinyu Ma** − *Suzhou MatSource Technology Co., Ltd., Suzhou 215000 Jiangsu, China;* orcid.org/0000-0001-5353-0652
**Rui Su** − *Suzhou MatSource Technology Co., Ltd., Suzhou 215000 Jiangsu, China*
**Gang Wang** − *College of Smart Materials and Future Energy, State Key Laboratory of Molecular Engineering of Polymers, Fudan University, Shanghai 200433, China*
**Muyao Yu** − *School of Physical Science and Technology, ShanghaiTech University, Shanghai 201210, P. R. China*
**Dong Zhong** − *Suzhou MatSource Technology Co., Ltd., Suzhou 215000 Jiangsu, China;* orcid.org/0009-0005-7534-8866
**Jie Zhu** − *The State Key Laboratory of Molecular Engineering of Polymers and Department of Macromolecular Science, Fudan University, Shanghai 200438, P. R. China*
**Weibo Gong** − *Gusu Laboratory of Materials, Suzhou 215000 Jiangsu, China*
**Huan Gu** − *Suzhou MatSource Technology Co., Ltd., Suzhou 215000 Jiangsu, China*
**Limin Li** − *Suzhou MatSource Technology Co., Ltd., Suzhou 215000 Jiangsu, China; Gusu Laboratory of Materials, Suzhou 215000 Jiangsu, China*
**Chen Shen** − *Gusu Laboratory of Materials, Suzhou 215000 Jiangsu, China*
**Rui Wu** − *Gusu Laboratory of Materials, Suzhou 215000 Jiangsu, China*
**Zhenghao Wu** − *Department of Chemistry and Materials Science, Xi'an J Liverpool University, Suzhou 215123 Jiangsu, P. R. China;* orcid.org/0000-0003-2862-4432
**Kan Xu** − *Suzhou MatSource Technology Co., Ltd., Suzhou 215000 Jiangsu, China*
**Min Zhou** − *Key Laboratory of Electroanalytical Chemistry, Changchun Institute of Applied Chemistry, Chinese Academy of Sciences, Changchun 130022, China;* orcid.org/0000-0003-4073-5370

Complete contact information is available at:
https://pubs.acs.org/10.1021/jacsau.6c01014

## Author Contributions

[⊥]C.M., L.Z., Y.C., W.D., S.F., Z. J., C.L., X.M., R.S., G.W., M.Y., D.Z., and J.Z. authors contributed equally to this work.

## Notes

The authors declare no competing financial interest.

## ■ ACKNOWLEDGMENTS

This work was supported by JSPS KAKENHI (Nos. JP25H01508 and JP25K01737), Suzhou MatSource Technology Co., Ltd. (Suzhou, China), and the Gusu Laboratory of Materials (grant number Y2501).

## ■ REFERENCES